\documentclass[a4paper,12pt, epsfig]{article}
\usepackage{epsfig}
\usepackage{amssymb,latexsym}
\usepackage{amsfonts}
\usepackage{amsmath}

\newskip\humongous \humongous=0pt plus 1000pt minus 1000pt

\newif\ifdtup

\catcode`\@=11

\@addtoreset{equation}{section}
\def\theequation{\thesection.\arabic{equation}}

\def\@normalsize{\@setsize\normalsize{15pt}\xiipt\@xiipt
\abovedisplayskip 14pt plus3pt minus3pt%
\belowdisplayskip \abovedisplayskip
\abovedisplayshortskip \z@ plus3pt%
\belowdisplayshortskip 7pt plus3.5pt minus0pt}

\def\small{\@setsize\small{13.6pt}\xipt\@xipt
\abovedisplayskip 13pt plus3pt minus3pt%
\belowdisplayskip \abovedisplayskip
\abovedisplayshortskip \z@ plus3pt%
\belowdisplayshortskip 7pt plus3.5pt minus0pt
\def\@listi{\parsep 4.5pt plus 2pt minus 1pt
      \itemsep \parsep
      \topsep 9pt plus 3pt minus 3pt}}

\relax

\catcode`@=12

\catcode`\@=11

\def\section{\@startsection{section}{1}{\z@}{3.5ex plus 1ex minus
    .2ex}{2.3ex plus .2ex}{\large\bf}}

\def\thesection{\arabic{section}}
\def\thesubsection{\arabic{section}.\arabic{subsection}}

\def\appendix{\setcounter{section}{0}
  \def\thesection{Appendix \Alph{section}}
  \def\thesubsection{\Alph{section}.\arabic{subsection}}
  \def\theequation{\Alph{section}.\arabic{equation}}}
\def\SymBoxes#1#2#3#4{\newdimen\un@t \un@t#3%
\raisebox{#1}{\rule{#2\un@t}{#4}\hskip-#2\un@t% lower horizontal
\@tempdimb\un@t \advance\@tempdimb by-#4\@tempcntb#2\relax%
\@whilenum{\@tempcntb>0}\do{%                         % #2 vertical lines
\rule{#4}{\un@t}\hskip\@tempdimb \advance\@tempcntb by\m@ne}%
\hskip-#2\un@t \rule[\un@t]{#2\un@t}{#4}%
\rule[\un@t]{#4}{#4}\hskip-#4%             % upper horizontal line
\rule{#4}{\un@t}}\hskip-#4}                % rightest vertical line
\begin{document}
%\begin{letter}{~}

%%%%%%Define some new commands and  macros

\newcommand{\dd}{\textrm{d}}

\newcommand{\beq}{\begin{equation}}
\newcommand{\eeq}{\end{equation}}
\newcommand{\bea}{\begin{eqnarray}}
\newcommand{\eea}{\end{eqnarray}}
\newcommand{\beas}{\begin{eqnarray*}}
\newcommand{\eeas}{\end{eqnarray*}}
\newcommand{\defi}{\stackrel{\rm def}{=}}
\newcommand{\non}{\nonumber}
\newcommand{\bquo}{\begin{quote}}
\newcommand{\enqu}{\end{quote}}
%%%%%%%%%%%%%%%%
\renewcommand{\(}{\begin{equation}}
\renewcommand{\)}{\end{equation}}
%%%%%%%%%%%%%%%%%%%%%%%%%%%%%%%%%% definitions
\def\de{\partial}
\def\Om{\ensuremath{\Omega}}
\def\Tr{ \hbox{\rm Tr}}
\def\rc{ \hbox{$r_{\rm c}$}}
\def\H{ \hbox{\rm H}}
\def\HE{ \hbox{$\rm H^{even}$}}
\def\HO{ \hbox{$\rm H^{odd}$}}
\def\HEO{ \hbox{$\rm H^{even/odd}$}}
\def\HOE{ \hbox{$\rm H^{odd/even}$}}
\def\HHO{ \hbox{$\rm H_H^{odd}$}}
\def\HHEO{ \hbox{$\rm H_H^{even/odd}$}}
\def\HHOE{ \hbox{$\rm H_H^{odd/even}$}}
\def\K{ \hbox{\rm K}}
\def\Im{ \hbox{\rm Im}}
\def\Ker{ \hbox{\rm Ker}}
\def\const{\hbox {\rm const.}}
\def\o{\over}
\def\im{\hbox{\rm Im}}
\def\re{\hbox{\rm Re}}
\def\bra{\langle}\def\ket{\rangle}
\def\Arg{\hbox {\rm Arg}}
\def\exo{\hbox {\rm exp}}
\def\diag{\hbox{\rm diag}}
\def\longvert{{\rule[-2mm]{0.1mm}{7mm}}\,}
\def\a{\alpha}
\def\b{\beta}
\def\e{\epsilon}
\def\l{\lambda}
\def\ol{{\overline{\lambda}}}
\def\ochi{{\overline{\chi}}}
\def\th{\theta}
\def\s{\sigma}
\def\oth{\overline{\theta}}
\def\ad{{\dot{\alpha}}}
\def\bd{{\dot{\beta}}}
\def\oD{\overline{D}}
\def\opsi{\overline{\psi}}
\def\dag{{}^{\dagger}}
\def\tq{{\widetilde q}}
\def\L{{\mathcal{L}}}
\def\p{{}^{\prime}}
\def\W{W}
\def\N{{\cal N}}
\def\hsp{,\hspace{.7cm}}
\def\bo{\ensuremath{\hat{b}_1}}
\def\bfo{\ensuremath{\hat{b}_4}}
\def\co{\ensuremath{\hat{c}_1}}
\def\cfo{\ensuremath{\hat{c}_4}}
\newcommand{\C}{\ensuremath{\mathbb C}}
\newcommand{\Z}{\ensuremath{\mathbb Z}}
\newcommand{\R}{\ensuremath{\mathbb R}}
\newcommand{\rp}{\ensuremath{\mathbb {RP}}}
\newcommand{\cp}{\ensuremath{\mathbb {CP}}}
\newcommand{\vac}{\ensuremath{|0\rangle}}
\newcommand{\vact}{\ensuremath{|00\rangle}                    }
\newcommand{\oc}{\ensuremath{\overline{c}}}
\newcommand{\Cos}{\textrm{cos}}

\newcommand{\Vol}{\textrm{Vol}}

\newcommand{\half}{\frac{1}{2}}

%%%%%%%%%%%%%%%%%%%%%%%Changed%%%%%%%%%%%%%%%%%%%%%%%%%%%%%
\def\changed#1{{\bf #1}}
%\def\changed#1{ #1}
%%%%%%%%%%%%%%%%%%%%%%%%%%%%%%%%%%%%%%%%%%%%%%%%%%%%%%%%%

\begin{titlepage}
%\begin{flushright}
%IFUP-TH/2011-??
%\end{flushright}
%\bigskip

\def\thefootnote{\fnsymbol{footnote}}

\begin{center}
{\large {\bf
Neutrino Dispersion Relations from a Dark Energy Model
  } }
%\end{center}

\bigskip

\bigskip

{\large \noindent Emilio
Ciuffoli$^{1}$\footnote{emilio.ciuffoli@gmail.com}, Jarah
Evslin$^{1}$\footnote{\texttt{jarah@ihep.ac.cn}}, Jie
Liu$^{2}$\footnote{liujie@ihep.ac.cn} and Xinmin Zhang$^{2,
1}$\footnote{\texttt{xmzhang@ihep.ac.cn}} }
\end{center}

\renewcommand{\thefootnote}{\arabic{footnote}}

\vskip.7cm

\begin{center}
\vspace{0em} {\em  { 1) TPCSF, IHEP, Chinese Acad. of Sciences\\
2) Theoretical physics division, IHEP, Chinese Acad. of Sciences\\
YuQuan Lu 19(B), Beijing 100049, China}}

\vskip .4cm

\end{center}

\vspace{4.5cm}

\noindent
\begin{center} {\bf Abstract} \end{center}

\noindent We consider a model in which the neutrino is kinetically coupled to a scalar field and study its
implications for environmentally dependent neutrino velocities.  Following the usual effective field theory logic, this coupling is expected to arise in neutrino dark energy models.  It leads to a Lorentz violation in the neutrino sector.  The coupling of the scalar field to the stress tensor
of the Earth yields terrestrial neutrino dispersion relations distinct from those in interstellar space.%, so as to be
%simultaneously compatible OPERA, MINOS and SN1987A data upon
%fitting a single parameter.

\vfill

\begin{flushleft}
{\today}
%\vspace{1cm}
\end{flushleft}
\end{titlepage}
%\bigskip

\hfill{}
%\bigskip

%\tableofcontents

\setcounter{footnote}{0}

\section{Introduction}
Last year the OPERA experiment announced the arrival of 16,111 Swiss neutrinos, 61 nanoseconds ahead of schedule \cite{opera}.  It turns out that this announcement was incorrect \cite{zichichi} and in the past year nearly all of the long baseline neutrino experiments \cite{tutti} including OPERA itself \cite{opera} have confirmed that neutrinos travel at the speed of light to within experimental errors.  But just as the velocity anomaly has disappeared, other anomalies and inconsistencies in the standard model with two or three mass neutrinos have been reinforced.  

The velocity anomaly at OPERA immediately was observed to be in contradiction with the measurement of the velocities of the neutrinos from SN1987A \cite{k21987,imb1987}.  There were attempts to evade this inconsistency with artificial models of energy-dependences \cite{strumia} but these were found to be in contradiction with simple kinematic arguments \cite{gg,xiaojun}.  The only models which could not be falsified on theoretical grounds alone were those which radially changed the nature of spacetime \cite{sformati} and those which introduced an environmental dependence in the neutrino mass \cite{alex}, although the later was found to be inconsistent with fifth force constraints \cite{fifth}, which was remedied in Ref.~\cite{tedeschi}.  In fact environmentally dependent neutrino models have been used to explain anomalies while avoiding various constraints for almost a decade \cite{nuenergy,wiener}.

Similarly there appears to be room to resolve today's anomalies using environmental dependence.  For example, the LSND experiment \cite{lsnd} and MiniBooNE \cite{miniboone} now both report an anomalous electron (anti)neutrino surplus which excludes the standard 3 neutrino oscillation paradigm at $3.8\sigma$.   Nearly all of their parameter is space is ruled out by the fact that no deficit was observed at KARMEN \cite{karmen} or ICARUS \cite{icarus}.  While the KARMEN and LSND experiments are nearly identical, the KARMEN baseline has a much lower average density than the LSND baseline.  In addition, inside of a gallium detector both the GALLEX \cite{gallex} and SAGE experiments \cite{sage} report a 2.5$\sigma$ deficit of neutrinos from a radioactive source.  In addition, tens of short baseline \cite{reattoreanom} and 1 km baseline \cite{noiunokm} reactor antineutrino experiments have found about a 6\% flux deficit.  Finally, at even higher densities, the expected upturn in solar neutrinos at low density has not been observed and there may even be evidence for a downturn \cite{smirnov}.  

One explanation for these anomalies could be sterile neutrinos, however combining ICARUS, KARMEN, LSND and MiniBooNE constraints one finds a sterile neutrino mass of about 0.5 eV \cite{icarus} while the gallium anomaly requires a mass about 0.8 eV or else there would be no oscillation inside of the 2 meter detectors and the solar anomaly requires a mass of a millielectron volt or less.  Three flavors of sterile neutrinos are strongly excluded by cosmological constraints.  While constraints arising from primordial helium abundance in HII regions are fraught with systematic errors and controversy, last years South Pole Telescope's measurement of the CMB diffusion radius \cite{spt} alone together with standard big bang nucleosynthesis and WMAP7's measurement of the sound horizon \cite{wmap7} provide a reasonably definitive exclusion of 3 sterile neutrino models.  Within the next year the new South Pole Telescope and Planck data will provide the last word on the number of flavors of light particles which were once in thermal equilibrium with the primordial plasma, providing a firm upper bound on the number of sterile neutrinos which are sufficiently strongly coupled to explain the anomalies listed in the previous paragraph.

Perhaps the correct explanation for these anomalies, like the neutrino velocity anomaly reported by OPERA, is experimental error.  However it is difficult to ignore the fact that they all occur in dense environments, such as the Earth, the iron shielding of LSND, gallium detectors and even the core of the sun.  The lowest density environment in which an anomaly has been reported is ice, where a horizontal neutrino excess has been reported by AMANDA and confirmed by IceCube \cite{icecube}.  However even in that case, this is a relative excess and could easily be instead a deficit of neutrinos which pass through the Earth at steeper angles.  The deficit saturates around 10-15 degrees, corresponding to a maximum baseline in the Earth and so could, for example, result from oscillation to a sterile neutrino which may or may not be more massive inside of the Earth.  

Summarizing, it remains a logical possibility that these anomalies are caused by some beyond the standard model matter effects.  After all, neutrinos are the only particles which can travel more than 1 km through solid rock, and so they are the only particles which may be sensitive to such new physics.

This motivates a systematic study of effective field theories encoding potential environmental effects on neutrinos.  While a large body of literature exists on beyond the standard model interactions of conventional matter on neutrinos, in the present work we will attempt a systematic study of a different class of models, which were first presented as explanations for the superluminality that had been claimed by OPERA.  These are models in which neutrinos interact not with the usual standard model particles, but with a dark coherent scalar field which is coupled gravitationally to other matter.  Such a dark scalar, if it has a fixed profile and fills a small portion of the universe, so that it does not increase in volume as it expands, will in recent times contribute to the density and pressure of the universe with a nonrelativistic equation of state and so will be a source of dark matter.  If on the other hand it nearly homogeneously fills most of the universe during some period, then it will contribute to the density and pressure with an equation of state $w=-1$ and so contribute to the dark energy during that period.  For simplicity and for a connection with the literature on this subject we will simply refer to the scalar field as the dark energy field, regardless of whether it ever had an equation of state close to -1.  In this note we will not attempt to use these models to solve the anomalies described above.

%This complimented a similar, but less confident, announcement concerning less energetic American neutrinos four years ago %\cite{minos}.
%Although the former of these announcements is claimed to be a 6 sigma event, as reviewed in Ref.~\cite{strumia} these %claims have been
%met with a degree of skepticism because they need to be reconciled with a series of yet more solid observations:
\newcounter{result}

\section{Neutrino Coupling and Dispersion Relations}

The models which we will consider invariably modify the neutrino dispersion relations.  This change necessarily violates Lorentz
symmetry.  We will consider spontaneous violations of Lorentz
symmetry, which arise by adding terms to the Lagrangian which
couple a new field to neutrino bilinears.  The couplings and the
new field are similar to those introduced in the neutrino dark
energy model of Ref.~\cite{nuenergy}, and to a large extent to
those in the earlier models of Refs.~\cite{nuenergyvec}, although
we do not demand that the additional field actually provides the
observed dark energy.  The field acquires a VEV due to
interactions with the Earth, which spontaneously breaks the Lorentz
symmetry.  In Refs.~\cite{alex,newerellis,grecia} such models were
constructed in which the new fields introduced were respectively a
symmetric tensor, a vector and a scalar.  We will consider a
scalar field $\Pi$, which as explained below will have the
advantage that it requires the tuning of only a single parameter.
The fact that our couplings resemble those which arise in the
neutrino dark energy scenario of Ref.~\cite{nuenergy} yields a
cosmological justification for the exclusiveness of these terms to
the neutrino sector.

In an effective field theory setting, it is sufficient to consider
the operators of lowest dimension which preserve
Lorentz-invariance.  The terms with no derivatives can be absorbed
into redefinitions of the fields and parameters of the effective theory.  Strong upper bounds on these terms arise, for example, on
masses from beta decay and as a result these terms will be
negligible at OPERA energies.  

As the standard model Higgs field has not yet been discovered\footnote{A new boson has been discovered, but the 6$\sigma$ confidence of the discovery arises from a $2\gamma$ excess which is greater than that of the standard model Higgs, and so cannot be considered to be evidence for the standard model Higgs.  The standard model Higgs does not yet fit the data $5\sigma$ better than the standard model with no Higgs, although it may before the scheduled shutdown.}  for the sake of generality we will not yet restrict our attention to a particular mechanism of SU(2) gauge symmetry breaking.  We will write the effective theory directly in terms of the neutrino fields, allowing the gauge symmetry to be realized either linearly or nonlinearly.  Later, in Eq.~(\ref{coupling}) we will specialize to the standard GWS symmetry breaking and we will see that this choice changes the naive dimension of our new operator, as factors of the Higgs field need to be added.  While the absolute dimension is indeed shifted, the relative dimensions of the different possible operators are not altered and therefore our interaction continues to be the dominant relevant interaction in the infrared.

We will be interested only in terms
which cannot be reabsorbed into other terms via field definitions
up to terms without derivatives.  In general the modifications of
the dispersion relation can be linear or quadratic in the new
couplings, we will eventually restrict our restriction to the
linear modifications and so to the coupling terms which lead to
linear modifications.  With all of these criteria, we are left
with \beq \Delta\mathcal{L}=\frac{1}{2}\left(
ia_\mu\bar{\nu}\partial^\mu\nu+ic_{\mu\nu}\bar{\nu}\gamma^\mu\partial^\nu\nu
-d_{\mu\nu\rho}\bar{\nu}\gamma^\mu\partial^\nu\partial^\rho\nu\right)
\eeq where $a$, $c$ and $d$ are tensors constructed from
derivatives of $\Pi$. In our dimensional analysis scheme, in which
the coefficients are constructed from a scalar field of dimension
$[m]$, the $d$ in the last term is of higher dimensionality and so
need not be considered, but we will keep it during this section
for illustration as a linear energy dependence of neutrino
superluminality is excluded at OPERA by less than two sigmas.

Clearly these derivative terms are nontrivial only if $\Pi$ is not constant.  The anomalies discussed above have been seen within the Earth and Sun, while SN1987A neutrinos tightly constrain the dispersion relation in space, therefore these derivatives need to be localized on Earth.  The simplest possibility would be if $\dot{\Pi}$ were localized on Earth and vanishes in space, while $\partial_k\Pi$ vanishes everywhere.  However it is easy to see that if $\Pi$ begins with a constant value, the spatial gradients between here and SN1987A will dominate over the temporal gradient in less than a mere 160,000 years, much less than the age of the Earth.  Therefore such field configurations are not logically consistent.

The next simplest possibility is that $\dot{\Pi}$ is negligible
%\footnote{It may be negligible for neutrino superluminality, but not for dark energy.}
 but the gradient of $\Pi$ lies along the Earth's radial direction $r$.  In Sec.~\ref{modsez} we will describe a prototypical example of a scalar field with the desired behavior.  For now, it will simply be relevant that the only nonzero derivatives, and so the only nonzero components of the tensors $a$, $c$ and $d$, will be those with only $r$ indices or an even number of identical tangential indices.  Then the dispersion relation is
\beq
E=\sqrt{P'^2+M'^2} \label{RelDisp}
\eeq
with $i$th spatial component
\beq
P'_i=p_i+c_{ij}p_j +d_{ijk}p_jp_k \qquad M'=m+a_jp_j
\eeq
where sums over repeated indices are understood.

Notice that as in the model of Ref.~\cite{newerellis} and unlike that of Ref.~\cite{alex}, the dispersion relation is anisotropic. 
% In particular, the velocity of a neutrino at the point $q$ traveling at an angle $\theta$ with respect to the radial direction is
 In particular, the velocity of a neutrino is 
\beq 
v_i=\frac{\partial E}{\partial p_i}=\frac{a_i M\p+(\delta_{ij}+c_{ij}+2d_{ijk}p_k)P'_j}{E} .
\eeq
Let $x$ be the radial coordinate and choose the $y$ direction to be orthogonal to $x$ such that the velocity of a given neutrino is in the $x-y$ plane, at an angle $\theta$ with respect to $x$.   Considering only the leading contributions to the Lorentz-violating terms, this neutrino will travel with a fractional superluminality equal to
\beq
\frac{v-c}{c}\simeq \frac{a_x^2}{2}\cos^2\theta+\left(c_{xx}\cos^2\theta+c_{yy}\sin^2\theta\right)
+2E\cos\theta\left(d_{xxx}\cos^2\theta+3d_{xyy}\sin^2\theta\right). \label{veq}
\eeq
We recall that, since the tensors $a$, $c$ and $d$ are proportional to the derivative of the scalar field $\Pi$, coefficients without an even number of identical tangential indices (such as $c_{xy}$ and $d_{yyy}$) are zero.

As mentioned above in the sequel we will ignore the last term as
it is created by a higher dimensional operator in our effective
field theory description.  Notice that $c_{xx}$ and $a_{x}$ only
occur in the combination $c_{xx}+a_{x}^2/2$ at this precision. The
terms which we have omitted are strongly suppressed, by factors of
%the superluminality fraction or even by the ratio of the neutrino energy to its rest mass
the Lorentz violating coefficients (which we assume to be at least of the same order of magnitude as the superluminality fraction) 
or even by the ratio of the neutrino rest mass to its energy, and so the identification of further
terms or even the separate identification of $c_{xx}$ and $a_{x}$
will need to await a much more precise or qualitatively different
experimental setting.  As we can then not hope to experimentally
distinguish between the effects of $c_{xx}$ and $a_x$, we will
simply neglect $a_x$ in what follows. Working with the effective
lagrangian with a  $SU_L(2) \times U_Y(1)$ symmetry,
% Due to the role of $W$ bosons in the mixing problem
%above, we will write the manifestly SU(2)-invariant form of
the second operator in Eq.(2.1) is given by 
\beq
\Delta\mathcal{L}=-b(\partial^\mu\partial^\nu \Pi) (H\dag\bar
L)\gamma_\mu\partial_\nu (LH) \label{coupling} \eeq where we have
defined the constant $b$ \beq c_{\mu\nu}=-\frac{b\langle
v\rangle^2}{2}\langle\partial_\mu\partial_\nu\pi  \rangle
\label{ceq} \eeq 
and the H is the Higgs doublet of the standard
model.
   In Ref.~\cite{grecia} the author considers a term which couples the neutrino kinetic
 tensor to $\partial_\mu\Pi\partial_\nu\Pi$.  However this term is of one energy dimension greater
 than (\ref{coupling}), and so is suppressed according to the usual logic of effective field theories.

%At this point one may already fit to the OPERA data to find $c$ on Earth.  Instead we will find it
%convenient to first describe our simple scalar model which determines $c$ in terms of the derivatives of a scalar
%field in Sec.~\ref{modsez}.  We will then fit this parameter to the OPERA result in Sec.~\ref{fit}.

\section{The Model} \label{modsez}

%A crucial difference between neutrinos observed by OPERA and those emitted by SN1987A is that the first traveled within the Earth, while most of the journey of the later was in interstellar space.  Thus the discrepancy  between these observations can be accounted for if the velocity of a neutrino is higher within the Earth than in space.  This can be arranged in a number of inequivalent ways, by 

For concreteness we will consider models in which the neutrino propagator is modified by a kinetic coupling of the neutrino to a field which obtains a VEV, if this messenger field is coupled to the Earth.  In Ref.~\cite{nano}, for example, the authors proposed that this spatial dependence can emerge in type IIB string theory model.

No such coincidence is required in models in which the messenger is coupled directly to baryon density or to the background stress tensor in such a way that it acquires a classical expectation value concentrated near massive objects.  So long as this classical field drops off sufficiently quickly from the sources, it will affect terrestrial neutrinos and not appreciably affect supernova neutrinos.  However it is important that the field drops off sufficiently quickly so that the Earth's effects dominate preferably over those of the Sun and certainly over those of the center of the galaxy.   In Ref.~\cite{alex} this was achieved by adding a spin two field whose inverse mass is fixed by hand to be roughly the inverse radius of the Earth (and necessarily less than the inverse distance to the Sun), while the coupling was chosen to yield the OPERA superluminality.  In this note we will present an alternative model in which the effects of the Sun and Galaxy are suppressed not by tuning another parameter, but simply by the derivative structure of our coupling to the neutrinos.

We consider a model with the neutrino dark energy term
$(\ref{coupling})$ coupling the neutrino to a scalar field with or
without a minimal Galileon coupling \cite{Galileon} given by the
boundary DGP model \cite{consistency} \beq
\L_\Pi=-\frac{1}{2}\partial_\mu\Pi\partial^\mu\Pi-\frac{a}{2}(\Box\Pi)(\partial_\nu\Pi)^2+4\sqrt{3\pi
G_N}\Pi T. \eeq The coefficient $a$, which {\it{can}} be taken to
be zero,  is a parameter of dimension $[l^3]$ which parameterizes
the nonlinear Galileon interaction.  $T$ is the trace of the
stress tensor of all of the matter, except for the scalar.  The
coupling of $\Pi$ to the stress tensor could in principle lead to
fifth forces beyond experimental bounds, as described in a very
similar setting in Ref.~\cite{fifth}, however a quick calculation
shows that only the product of the coefficient of this coupling
and the coefficient $b$ in (\ref{coupling}) appears in the
neutrino $v-c$, therefore any reduction of the coefficient of the
stress tensor coupling which may be mandated by fifth force
constraints can be compensated by an opposite rescaling of $b$.
While the Lagrangian itself has higher derivative terms, terms in
the equations of motion have at most two derivatives acting on
each $\Pi$, which allows the existence of ghost-free solutions
such as that which we will use.

The Galileon interaction term is useful because it reduces short
distance singularities, via the Vainshtein mechanism
\cite{Vainshtein}, at least in the presence of spherically
symmetric stress tensor sources.  More precisely, for an external
source of mass $M$ there will be a distance scale
\cite{consistency} \beq R=\frac{(\pi
G_N)^{1/6}}{\sqrt{3}}\left(4aM\right)^{1/3} \eeq at which the
behavior of the $\Pi$ field changes.  We can choose the Galileon
coupling $a$ such that this distance is either larger than or
smaller than the radius of the Earth $r$.  For concreteness, in
the rest of this note, we will chose $a$ to be sufficiently small
so that $R<<r$, and so we will set $a=0$ in the rest of this
section.  A small value of $a$ is useful for, among other things,
avoiding the potential formation of closed timelike curves
\cite{taotao}.  However $a$ will again become important for
distance scales smaller than $R$, for example it may be invoked
for neutrino phenomenology in neutron stars and, depending on its
value, in the cores of massive stars.

Let us now calculate a static, spherically symmetric field configuration $\Pi(r)$  in the presence of a nonrelativistic matter source with density $\rho(r)$.  The $\Pi$ equation of motion is then
\beq
-4\sqrt{3\pi G_N}\rho=\nabla^2\Pi=\frac{1}{r^2}\partial_r(r^2\partial_r\Pi). \label{edm}
\eeq
This is easily integrated to yield
\beq
r^2\partial_r\Pi=-4\sqrt{3\pi G_N}\int_{\tilde{r}=0}^{\tilde{r}=r}\tilde{r}^2\rho d\tilde{r}=-\sqrt{\frac{3 G_N}{\pi}} M(r) \label{dp}
\eeq
where $M(r)$ is the density contained in the object up to a radius $r$, and we have fixed the constant
of integration by imposing that $\Pi$ be differentiable at the origin.

One can now easily find the second derivatives of $\Pi$, which appear in the coupling (\ref{coupling}).   Choosing our coordinates such that $y=z=0$ at a given point
\bea
\partial_z^2\Pi&=&\partial_y^2\Pi=\frac{1}{r}\partial_r\Pi=-\sqrt{\frac{3 G_N}{\pi}}\frac{M(r)}{r^3}\label{ddz}\\
\partial_x^2\Pi&=&\nabla^2\Pi-\partial_y^2\Pi-\partial_z^2\Pi=-4\sqrt{3\pi G_N}\left(\rho(r)-\frac{M(r)}{2\pi r^3}\right)\label{ddx}
\eea
while terms with mixed derivatives vanish.  In particular, in the Earth's crust
\beq
\partial_x^2\Pi=4\sqrt{3\pi G_N}(\frac{2}{3}\rho_0-\rho)\hsp
\partial_y^2\Pi=\partial_z^2\Pi=-4\sqrt{\frac{\pi G_N}{3}}\rho_0 \label{buccia}
\eeq
where $\rho$ and $\rho_0$ are respectively the average densities of the crust and of the Earth as a whole.  More generally, Eq.~(\ref{buccia}) may be applied to a point at any radius $r$ in a background with an arbitrary spherically symmetric density profile if one identifies $\rho$ with the density at radius $r$ and $\rho_0$ with the average density at radii less than $r$.

%In the Galileon case $a>0$, when coupled to a static point source of mass $M$ the $\Pi$ field acquires a nonlinear profile %which plays the same role of that of Vainshtein's solution of massive gravity in Ref.~\cite{Vainshtein}.  The characteristic %radius of this solution is \cite{consistency}
%\beq
%R=\frac{(\pi G_N)^{1/6}}{\sqrt{3}}\left(4aM\right)^{1/3}
%\eeq
%and the solution at a displacement of $\vec x$ with respect to the core is described by
%\beq
%\partial_i \Pi(r)=\frac{9\sqrt{6}}{4a}\left(-1+ \sqrt{1+R^3/(18\pi r^3)}\right)x_i
%\eeq
%where we have restricted our attention to the branch which remains finite as $a\to 0$.
%If $r>>R$, which is automatic in the case $a=0$, this becomes
%\beq
%\partial_i \Pi(r)=\frac{\sqrt{6}R^3x_i}{16\pi ar^3}
%\eeq
%and so the second derivative which enters in the coupling (\ref{coupling}) is
%\beq
%\partial_r^2\Pi(r)=-\frac{\sqrt{6}R^3}{8\pi ar^3}=-\frac{\sqrt{G_N}}{4\sqrt{2\pi} }\frac{M}{r^3}. \label{epsilon}
%\eeq

\section{Fitting}  \label{fit}

\subsection{SN1987A}

The equation (\ref{coupling}) modifies the dispersion relation for the neutrinos, for example allowing their velocity to deviate from the usual relativistic form.  In the language of Ref.~\cite{alex} this corresponds to a modification of the effective metric in which the neutrinos propagate.  The neutrino velocity at an angle $\theta$ with respect to the radial direction is given by inserting Eq.~(\ref{ceq}) into (\ref{veq})
\beq
\epsilon=\frac{v-c}{c}=-\frac{b\langle v\rangle^2}{2}(\partial_x^2\Pi(r)\cos^2(\theta)+\partial_y^2\Pi(r)\sin^2(\theta)), %=\frac{b\sqrt{G_N}\langle v\rangle^2}{4\sqrt{2\pi} }\frac{M_E}{r_E^3}
\eeq
where $\langle v\rangle$ is the Higgs VEV.   In the case of even the longest baseline neutrino experiments to date, $\cos^2(\theta)\sim .01$ and so we will ignore the first term.
Using Eq.~(\ref{buccia}) one finds
\beq
\epsilon=\frac{v-c}{c}=2b\sqrt{\frac{\pi G_N}{3}}\langle v\rangle^2\rho_0\sim(800\ eV)^5 b\label{radvel}
\eeq
where we have used $\rho_0\sim 5.5$ gm/cm${}^3$.  For example the OPERA result $\epsilon\sim 3\times 10^{-5}$ would be reproduced if
\beq
b\sim\frac{1}{(6\ {\rm{keV}})^5}.
\eeq
This is much smaller than the energies of the OPERA neutrinos, which means that this effective field theory approach is invalid for deviations from relativistic dispersion relations as large as those reported by OPERA.  The effective field theory approach can only be trusted below the cutoff energy.   

Of course, this fit is only reasonable if the main source of the dark energy field is the Earth.  In other words, the main contribution to $\Pi$ must arise from its coupling to the Earth, and not for example the Sun or the matter in our galaxy.  In the case of a distance object, the $\rho$ term in Eq.~(\ref{ddx}) vanishes, while the other terms are simply the ratio of the total mass to the distance cubed.  Therefore, at the surface of the Earth, the contribution of the Sun (with respect to that of the Earth) is suppressed by a factor of $10^{-8}$ and that of the galaxy by $10^{-25}$, thus it is reasonable to ignore these contributions.  In the case of the model in Ref.~\cite{alex} the corresponding $M/R$ dependence would lead to a dominant contribution from the Sun and galaxy, and so these contributions are eliminated by hand by choosing the mass of their new field to be of order the inverse radius of the Earth.

And so what about neutrinos from SN1987A?  The fractional superluminality for these neutrinos needs to be less than $10^{-9}$.  This is easily satisfied,
as one can see by directly calculating the change in arrival time, or else simply noting that nearly all of the trip is in interstellar space,
where $\epsilon$ is of order $10^{-30}$.  In fact, examining Eq.~(\ref{ddx}) more closely, one sees that during the first part of their trip the
neutrinos are actually subluminal.  The subluminality constraints on SN1987A neutrinos are very weak, as they depend upon assumptions about the propagation of light in the various media between here and the large Magellanic cloud.

\subsection{KamLAND} \label{kamsez}
%Fitting the OPERA and MINOS results is easy, one need merely add a Lorentz-violating term to the muon neutrino sector.  The SN1987A neutrinos are then not a problem, so long as the electron neutrino sector is unaltered.  This is also consistent with LEP synchrotron limits, as these concern electrons, if one ignores flavor mixings.  The limits on muon Lorentz violation placed by gamma rays arriving from blazars \cite{muon} are violated, but since our scalar is only appreciable during a small portion of this trip, the relevance of these bounds to OPERA physics is unclear.  However as we have alluded earlier, the biggest problem arises from KamLAND's reactor neutrino oscillation data.  This suggests that certain terms which violate Lorentz-invariance in the muon neutrino sector must be equal in the electron neutrino sector.  No such analysis has been carried out, however, for every possible term in our low energy effective description.  In this subsection we will comment on the possibility of evading these bounds within the validity of the effective theory.

KamLAND lies between 55 nuclear reactors, which supply it with antineutrinos.  The electron neutrinos oscillate into muon neutrinos.  If the energy difference between an electron and muon neutrino is $\Delta E$, the probability that a given electron neutrino is still an electron neutrino after traveling a distance $L$ is
\beq
1-\sin^2(2\theta_{e\mu})\sin^2\left(\frac{L\Delta E}{2}\right)
\eeq
where $\theta_{e\mu}$ is the mixing angle between these two flavors.  Using the dispersion relation (\ref{RelDisp}) one readily finds, for an energy $E$ neutrino, that the latter phase is
\bea
&&L\left(\frac{\delta m^2}{2E}+\left(a_{z}^{(e)2}+2c_{zz}^{(e)}-a_{z}^{(\mu)2}-2c_{zz}^{(\mu)}\right)E \cos^2\theta+ 2\left(c_{yy}^{(e)}-c_{yy}^{(\mu)} \right)E\sin^2\theta +\right. \nonumber\\ &&
\phantom{\frac{\delta m^2}{2E}}\left.\left(2\left(d_{xxx}^{(e)}-d_{xxx}^{(\mu)} \right)\cos^2\theta+
3\left(d_{xyy}^{(e)}-d_{xyy}^{(\mu)} \right)\sin^2\theta\right)E^2\cos^2\theta\right)
\eea
where $\delta m^2$ is the difference between the squared masses of the two relevant neutrino mass eigenstates and the $e$ and $\mu$ superscripts
denote the Lorentz-violating neutrino couplings to the two flavors respectively.

This can be easily applied to models that had been proposed for the neutrino velocity anomaly.  Inserting coefficients $2c+a^2$ which are sufficiently large
to fit the OPERA data, together with an energy $E=5$ MeV typical of reactor neutrinos seen above the background at KamLAND, one finds that neutrinos
oscillate about every 5 nanometers.  This is much smaller than the resolution at KamLAND, and so KamLAND would in this case  find an energy-independent
survival probability.  This is in grave contradiction with KamLAND's results in Ref.~\cite{kamland} which present two full averaged neutrino oscillations.
Higher order terms, such as the $d$ term, lead to an even shorter wavelength but with an $E$ dependence which cannot cancel that of previous terms.
Therefore a cancellation in such terms, in order to be consistent with KamLAND's results, must be imposed order by order in $E$, and in fact also order
in order in $\cos(\theta)$, which is small but still larger than the ratio of the energy to the neutrino mass.  Such a cancellation is not possible
with the terms in the lowest orders of the effective action considered in this note.

%\section{Observational Signatures and Alternative Couplings}

\subsection{Solar Neutrinos and the MSW effect}

Our model modifies the neutrino dispersion relations not only near the Earth, but near any massive body.  In fact the effect is even more pronounced near the Sun, although unlike the model of Ref.~\cite{alex} the modification of the fractional neutrino velocity in the core of the Sun is still quite small.  This modification means that experimental signatures for our model may already be apparent in the solar neutrino data, and in particular solar neutrino data provides a nontrivial experimental check on the viability of our proposal.  The velocity of solar neutrinos cannot be determined directly, as one does not know the time at which they were emitted.  However the interaction of neutrinos with the Sun leads to a further modification of their dispersion relations which has an observed effect on neutrino oscillation, called the MSW effect \cite{msw}.  This leads one to ask whether the modification to the neutrino dispersion relation which we propose modifies the experimental signatures of the MSW effect, in particular whether it modifies the energy dependence of the electron neutrino survival probability.

In this subsection we will modify a textbook \cite{strumiarev} derivation of the solar electron neutrino survival probability, applying the superluminal dispersion relation of our model.  We will calculate the change in the survival probability for neutrinos of low enough energy such that their flavors are converted adiabatically, as is the case for observed solar neutrinos.  While we will find that this effect is much too small to be observed for solar neutrinos, we will note that it is potentially large enough to be observed some day for supernova neutrinos.

The relevant electroweak interaction of neutrinos with leptons and baryons can be described by the Lagrangian density term of the form
\(
\Delta \mathcal{L}=\bar{\nu} A \gamma_0 P_L \nu .
\)
If we consider only two flavors (as is reasonable for rough calculations involving solar neutrinos) we can express the interaction parameter $A$ as
\(
A=\sqrt{2}G_F\left[\frac{N_e}{2}\left(\begin{array}{cc} 1 & 0\\ 0&-1\end{array}\right)-\frac{N_n-N_e}{2}\left(\begin{array}{cc} 1 & 0\\ 0&1\end{array}\right)\right] 
\)
where $N_e$ and $N_n$ are the electron and neutron densities respectively.  We consider, for simplicity, neutrinos which travel radially outwards from the center of the Sun, in the direction which we will call $\hat{x}$. Recall from Eq.~(\ref{buccia}) that the main contribution to the superluminality fraction $\epsilon$ arises from
\(
c_{xx}^\odot=-2b\langle v^2\rangle\sqrt{3\pi G_N}\left(\frac{2}{3}\rho_0^\odot-\rho^\odot\right)
\)
where $c_{xx}^\odot$ is value of the Lorentz-violating (LV) coefficient defined in Eq. (\ref{ceq}) inside of the Sun.  $\rho^\odot(r)$ and  $\rho_0^\odot(r)$ are the solar density at radius $r$ and the mean density at radius less than $r$ respectively.  The $r$ dependence will usually be left implicit.

In the case of the OPERA experiment, since the neutrino velocity is nearly perpendicular to the radial direction, the main contribution to the superluminality was proportional to the value of the LV coefficient inside of the Earth $c_{yy}^{\oplus}=2b\langle v^2\rangle\sqrt{\pi G_N} \rho_0^\oplus/\sqrt{3}$.

In the center of the Sun, $\rho_0^\odot(0)=\rho^\odot(0)$, and hence $c_{xx}^\odot$ and $c_{yy}^\odot$ are equal
\(
\left(\frac{2}{3}\rho_0(0)-\rho(0)\right)= -\frac{1}{3}\rho_C^\odot
\quad \Rightarrow \quad 
c_{xx}^\odot(0)=2b\langle v^2\rangle\sqrt{\frac{\pi G_N}{3}}\rho^\odot_C=\frac{\rho^\odot_C}{\rho_0^\oplus}\epsilon^\oplus \simeq 8\cdot 10^{-4} 
\)
where $\epsilon^\oplus$ is the superluminality inside of the Earth's crust (for tangential motion), $\rho^{\odot}_C$ is the density in the center of the Sun and $\rho^{\oplus}_0$ is the average density of the Earth. 

In order to calculate $\rho_0^\odot$ at a generic point, we need the approximate density profile of the solar core
\(
\rho^\odot(r)\simeq \rho^\odot_C\exp\left[-\frac{10 r}{R_\odot}\right] .
\)
On the boundary of solar core ($r=0.2 R_\odot$)
\(
\left(\frac{2}{3}\rho_0(0.2R_\odot)-\rho(0.2R_\odot)\right)\simeq \frac{1}{40}\rho^\odot_C \quad \Rightarrow \quad 
c_{xx}^\odot(0.2R_\odot)=-\frac{3\rho^\odot_C}{40\rho_0^\oplus}\epsilon^\oplus \simeq -0.6\cdot 10^{-4}.
\)
This means that at the center of the Sun the neutrinos are superluminal but at higher radii they slow and by the boundary of the core they are subluminal. Crucially, at every radius the dimensionless LV parameter $c_{xx}^\odot$ is much smaller than one, and so we can use it to perturbatively expand all quantities of interest.

Including the interaction term $A$, in the ultrarelativistic limit $E>>m$ our dispersion relation (\ref{RelDisp}) becomes
\(
E_{LV}\simeq(1+c_{xx}^\odot)p +\left(\frac{m \cdot m^\dagger}{2E_{LV}}+A\right) \label{esole}
\)
where we have defined the mass matrix
\(
m\cdot m^\dagger=V\left(\begin{array}{cc} m_1^2 & 0 \\ 0 & m_2^2 \end{array} \right)V^T, \qquad
 V=\left(\begin{array}{cc}\Cos\theta & \sin\theta \\ -\sin\theta & \Cos\theta \end{array} \right) .
\)
The subscript $LV$ indicates that the dispersion relation is Lorentz-violating. The evolution of a neutrino planewave wavefunction in the electroweak interaction basis is given by multiplication by
\beq
e^{-ipx+iE_{LV} t}=e^{(-ipv+iE_{LV})t}=e^{(-ip(1+c_{xx}^\odot)+iE_{LV})t}=e^{iHt}
\eeq
where we have defined the matrix $H$, which can be evaluated using Eq.~(\ref{esole}) 
\(
H=\frac{m\cdot m^\dagger}{2E_{LV}}+A .
\) 
The matrix H has the same form as in the Lorentz-invariant (LI) case, but now $E_{LV}\simeq (1+c_{xx}^\odot)p$ instead of $E\simeq p$.% (recall that our LV coefficients $c_{\mu\nu}$ are flavor independent).

In our two-flavor approximation, the mixing angle and the difference between the two eigenvalues can be found diagonalizing the matrix H
\(
H=\frac{m_1^2+m_2^2+2\sqrt{2}G_FE_{LV}(N_e-N_n)}{4E_{LV}}\left(\begin{array}{cc} 1 & 0 \\ 0& 1\end{array} \right)+ \frac{\Delta m^2}{4E_{LV}}\left(\begin{array}{cc}-\Cos2\theta +\delta_{LV} & \sin2\theta \\ -\sin2\theta & \Cos2\theta-\delta_{LV} \end{array} \right)
\)
where
\(
\delta_{LV}=\frac{2\sqrt{2}G_FE_{LV} N_e}{\Delta m^2}\simeq7.6\cdot 10^{-8}
\frac{E_{LV}/{\textrm{MeV}}}{\Delta m^2/{\textrm{eV}^2} }
\frac{\rho}{\textrm{g/cm}^3}Y_e
\)
and $Y_e=Z/A$ is the electron fraction. The terms proportional to the identity matrix affect neither the difference $\Delta m_m^2$ between the eigenvalues of $H$ nor the rotation angle $2\theta_m$ which diagonalizes~it
\bea \label{parOsc}
\left.\Delta m_m^2\right|_{LV}&=&\Delta m^2\sqrt{(\Cos(2\theta)-\delta_{LV})^2+\sin^2(2\theta)}, \\ \label{parOsc2} \left.\sin(2\theta_m)^2\right|_{LV}&=&\frac{\sin(2\theta)}{\sqrt{(\Cos(2\theta)-\delta_{LV})^2-\sin^2(2\theta)}}.
\eea
%Our LV coefficient $c_{xx}^\odot$ is small: the ratio between the density in the core of the Sun and in the average density of the Earth is in the range
%\( 4<\frac{\rho^{\odot}}{\rho^{\oplus}_0} < 30. \)
%As our LV parameters grow linearly with the density, in the core of the Sun (and in the radial direction) we have $5\cdot 10^{-5} <c_{xx}^\odot< 3\cdot 10^{-4}$.

Expanding in $c_{xx}^\odot$ , Eqs.~(\ref{parOsc}) and (\ref{parOsc2}) yield
\bea
 \left.\Delta m_m^2\right|_{LV}\simeq&& \Delta m_m^2\left(1-\frac{\delta(\Cos(2\theta)-\delta) }{\lambda}c_{xx}^\odot\right)
\nonumber \\
\left.\sin^2(2\theta_m)\right|_{LV}\simeq &&\sin^2(2\theta_m)\left(1+\frac{\delta(\Cos(2\theta)-\delta) }{\lambda}c_{xx}^\odot\right)
\eea
where 
\(
\lambda=(\Cos2\theta-\delta)^2+\sin^22\theta 
 \)
and $\delta$ is obtained from $\delta_{LV}$ by replacing the Lorentz-violating dispersion relation with the usual one
\(
\delta=\frac{2\sqrt{2}G_FE N_e}{\Delta m^2} .
\)

A typical value of the LV coefficient $c_{xx}$ inside of the core is
\[
\frac{1}{0.2R_\odot}\int_{0}^{0.2R_\odot}c_{xx}(r)\textrm{d}r\simeq -\frac{1}{4}\frac{\rho_C^\odot}{\rho^\oplus}\epsilon^\oplus\simeq -2\cdot 10^{-4}.
\]
%Considering a LV coefficient $c_{xx}^\odot\simeq 10^{-4}$, 
Taking $\Delta m^2\simeq 10^{-5} \textrm{eV}^2$, $\sin(2\theta)\simeq 0.9$, using the typical values for the solar core $\rho^\odot \simeq 80 \textrm{g/cm}^3$, $Y_e\simeq 1$ and $|p|\simeq 1 \textrm{ MeV}$, the fractional difference between the Lorentz-violating and the Lorentz-invariant effective mass squared difference is
\(
\frac{\left.\Delta m_m^2\right|_{LV}-\Delta m_m^2}{\Delta m_m^2}\simeq
-\frac{\left.\sin(2\theta_m)^2\right|_{LV}-\sin(2\theta_m)^2}{\sin(2\theta_m)^2}\simeq 4\cdot 10^{-6}
\)
which is much smaller than the fractional uncertainties in the values of $\Delta m^2$ and $\sin(2\theta_m)^2$.

Moreover, since the solar radius is much larger than the oscillation length, the surviving probability of an electron neutrino detected on the Earth is (neglecting the effect of the matter in the Earth)
\(
P_{(\nu_e\rightarrow\nu_e)}=\frac{1}{2}+\frac{1}{2}\Cos(2\theta)\Cos(2\theta_m) .
\)
In our LV case we find
\bea
\left.P_{(\nu_e\rightarrow\nu_e)}\right|_{LV}&=&P_{(\nu_e\rightarrow\nu_e)}\left(1-\frac{(\Cos(2\theta)-\delta)\Cos(2\theta)\sin^2 (2\theta) \delta }{\lambda\left(\sqrt{\lambda-\sin^2(2\theta)}(\sqrt{\lambda}+\Cos(2\theta)\sqrt{\lambda-\sin^2(2\theta)}\right)} c_{xx}\right)\nonumber\\
&\simeq &P_{(\nu_e\rightarrow\nu_e)}(1-0.1\cdot c_{xx}) .
\eea
In the case of solar neutrinos $c_{xx}=c_{xx}^\odot\sim 10^{-4}$ and so we obtain a correction to the neutrino survival probability of order $10^{-5}$, which is well within the experimental errors of solar neutrino observations and also within the uncertainties within which the relevant parameters are known.  Thus no deviation with respect to solar neutrino experiments is expected.  

If, on the other hand, the MSW effect is observed for supernova neutrinos then the adiabatic conversion assumed here is unreliable.  However the high density involved leads to a $c_{xx}$ of at least order unity, and so the corresponding correction to the conversion rate in this model will be appreciable, perhaps allowing for the exclusion of the model.

%We have
%\( \frac{\left.\Delta m_m^2\right|_{LV}-\Delta m_m^2}{\Delta m_m^2}=\frac{\delta(\Cos(2\theta)-\delta)c_{xx}}{\left(\Cos(2\theta)-\delta\right)^2+\sin2\theta^2}\simeq \)

\section{Experimental Signatures}

Once one has included the interaction term (\ref{coupling}) to explain superluminal neutrino velocities here on Earth, one needs to deal with its consequences throughout our universe.  In particular, it may become appreciable inside of various astrophysical objects.  In the case of a spherically symmetric object, Eq.~(\ref{buccia})  tells us that the superluminality at a radius $r$ only depends on the density $\rho$ at that radius and the average density at lower radii $\rho_0$.  This may seem peculiar, since there is no Gauss' law for $\Pi$.  Indeed, any choice of boundary conditions for $\Pi$ implies that its value in the core of an object depends on the $\rho$ profile at higher radii.  However in our model, and not in that of Refs.~\cite{alex,newerellis}, the superluminality only depends on the second derivatives of $\Pi$, which is independent of the constants of integration and so of these boundary conditions.  This freedom in choosing boundary conditions is necessary for many cosmological applications of the $\Pi$ field, such as the dark energy model of Ref.~$\cite{nuenergy}$ and Galileon cosmologies.

In the core of a spherically symmetric object, in which the density is approximately constant, the deviation from the relativistic dispersion relations is easily approximated.  Up to geometrical factors whose magnitude and sign depend on the direction in which the neutrino travels, the fractional  superluminosity of neutrinos is simply given by the ratio of the deviation on Earth to the ratio of the average density of the Earth to that of the object's core.  For example in the case of the core of our Sun, whose density varies between 4 and 30 times that of the Earth, the OPERA velocity anomaly would predict a neutrino fractional subluminosity or superluminosity of order between $10^{-4}$ and $10^{-3}$.  This change in the neutrino's dispersion relation affects its propagator and so may in principle have observable consequences on the various fusion processes occurring in our solar core, perhaps allowing for an exclusion or verification of our model. This is in contrast with the model of Ref.~\cite{alex}, in which neutrino superluminality is of order $1$ at the surface of the Sun and persists at greater depths including the outer core, where a significant amount of fusion occurs.

However, even in our model the superluminality consistency with OPERA's claim would have led to order 1 deviations from relativity whenever the density reaches about $2\times 10^5$ $g/cm^3$.  This level is easily surpassed for example in the cores of older massive stars which fuse carbon or heavier elements Ref.~\cite{stars}, and is even reached in white dwarves.  Thus in these situations one may expect drastic departures from the standard Lorentz-invariant quantum field theory predictions for amplitudes and decay rates involving the neutrino propagator, potentially affecting for example models of the helium flash.  This fact also implies that such models are in general incompatible with anomalies as large as those reported by OPERA.

Another crucial distinction between our model and that of Ref.~\cite{alex} is that our neutrino velocity is direction-dependent,
radially traveling neutrinos on the surface of the Earth are subluminal.  For shallow angles such as those of current accelerator neutrino experiments, the deviation from the relativistic dispersion relations in our model is reasonably independent of the angle, there is only a correction of order the angle
 $(\frac{\pi}{2}-\theta)$ squared.  On the other hand, in the second model of Ref.~\cite{newerellis}, the superluminality is
 proportional to the cosine of the angle with respect to a preferred direction, which given the symmetries of the problem
 will likely correspond to the Earth's radial direction.  

%As the baseline of T2K is half that of the others, this cosine
 %is only one half of its value in OPERA and MINOS, and so the second model of Ref.~\cite{newerellis} suggests that the
% neutrino $(v-c)/c$ at T2K is about one half its value at OPERA, corresponding to a neutrino arrival time only 14 nanoseconds
% ahead of schedule.  Thus the fact that T2K has observed luminal neutrinos in fact on its own does not rule out that model.  On the other hand, to distinguish our model from that of Ref.~\cite{alex} one needs a significantly larger angle, translating into a larger  baseline.  Almost an order of magnitude of improvement in the timing precision of OPERA (at both experiments) would be necessary to distinguish the models by comparing superluminality at OPERA and at Fermilab's proposed 1300 km baseline experiment LBNE.  The prospects are improved significantly in the 2000+ km proposed experiments sending neutrinos from CERN to the Pyh\"asalmi mine.

\section{Electron Superluminality?}
In addition to the experimental constraints described above one may add that bounds on terrestrial electron superluminality are extremely tight.  At LEP \cite{LEP}, synchrotron
radiation measurements at GeV energies yield a maximum fractional superluminality $(v-c)/c$ of $5\times10^{-15}$.
%\item SU(2) gauge-invariance requires that above the electroweak breaking scale, electrons and neutrinos travel at the same velocity.
%\end{list}
%Together with constaint (2),
This implies that either the neutrino
superluminality disappears before the electroweak scale, or else
some conspiracy of other effects prevents it from being
transmitted to the electron.  %The first scenario is very difficult
%to realize, especially in light of the fact that OPERA neutrino
%energies are already so close to the electroweak scale.  For
%example i

In Ref.~\cite{strumia} the authors noted, as was shown in
Ref.~\cite{nuenergyvec}, that in the case of a coupling
$\nu\dag\partial_0\nu$, a one-loop mixing generates the operator
$e\dag\partial_0 e$ which leads to a superluminality for electrons
which is suppressed by at most $E^2/M_W^2$ with respect to that of
neutrinos, and so is 5 orders of magnitude above the LEP bound
quoted above.  More complicated breaking models, such as the
coupling to sterile neutrinos \cite{nuenergyvec,strumia} and
supersymmetric completions \cite{strumia}, are able to push these
effects onto higher derivative operators.  However generically
this merely increases the exponent of the $E/M_W$ suppression.%, but
%since $E/M_W$ is of order $0.1$ it seems difficult to see how such
%models could be consistent with the LEP constraint.

One might hope that there exist flavor-asymmetric Lorentz-violating terms which evade the neutrino oscillation constraints of observation (2).   In this case one could simply declare that Lorentz-violation only occurs for muon and perhaps tau neutrinos, hoping that their mixing with electrons is sufficiently suppressed to avoid the LEP synchrotron bound.    There are a number of reasons why this approach fails, but we have seen (Subsec.~\ref{kamsez}) that our terms certainly do not allow such an evasion of the flavor neutrality constraints.

\section* {Acknowledgement}

\noindent JE is supported by the Chinese Academy of Sciences
Fellowship for Young International Scientists grant number
2010Y2JA01. EC, JL and XZ are supported in part by the NSF of
China.

%%%%%%%%%%%%%%%%%%%%%%%%%%%%%%%%

\end{document}

\bibitem{muon}
  B.~Altschul,
  ``Astrophysical limits on Lorentz violation for all charged species,''
  Astropart.\ Phys.\  {\bf 28 } (2007)  380-384.
  [hep-ph/0610324].

\bibitem{ellis}
  J.~R.~Ellis, N.~E.~Mavromatos, D.~V.~Nanopoulos and A.~S.~Sakharov,
  ``Space-time foam may violate the principle of equivalence,''
  Int.\ J.\ Mod.\ Phys.\  A {\bf 19} (2004) 4413
  [arXiv:gr-qc/0312044].
  %%CITATION = IMPAE,A19,4413;%%

\bibitem{newellis}
  G.~Cacciapaglia, A.~Deandrea and L.~Panizzi,
  ``Superluminal neutrinos in long baseline experiments and SN1987a,''
  arXiv:1109.4980 [hep-ph].

\bibitem{noidic}
  E.~Ciuffoli, J.~Evslin, X.~Bi and X.~Zhang,
  ``Density-Dependent Neutrino Dispersion Relations for OPERA?,''
  arXiv:1112.3551 [hep-ph].

\end{thebibliography}


\begin{thebibliography}{23}\setlength{\itemsep}{-2.3mm}

%%%%%%%%%%%%%%%%%%%%%%%%%%%%%%%%%

\bibitem{opera}
  T.~Adam {\it et al.} [ OPERA Collaboration ],
  ``Measurement of the neutrino velocity with the OPERA detector in the CNGS beam,''
[arXiv:1109.4897 [hep-ex]].

\bibitem{zichichi}
A.~Zichichi,
``Results from LVD-OPERA Combined Analysis: A Time-Shift in the OPERA Setup,"
available online at http://agenda.infn.it/getFile.py/access?resId=0\&materialId=slides\&confId=4896.
 N.~Y.~.Agafonova {\it et al.}  [LVD and OPERA Collaborations],
  ``Determination of a time-shift in the OPERA set-up using high energy horizontal muons in the LVD and OPERA detectors,''
  Eur.\ Phys.\ J.\ Plus {\bf 127} (2012) 71
  [arXiv:1206.2488 [hep-ex]].
 
\bibitem{tutti}
P. Adamson,
``Neutrino Velocity: Results and prospects of experiments at other beamlines,"
Talk presented at Neutrino 2012, Available Online at http://neu2012.kek.jp/neu2012/programme.html.
  P.~Alvarez Sanchez {\it et al.}  [Borexino Collaboration],
  ``Measurement of CNGS muon neutrino speed with Borexino,''
  Phys.\ Lett.\ B {\bf 716} (2012) 401
  [arXiv:1207.6860 [hep-ex]].
F.~Ronga,
  ``Analysis of the MACRO experiment data to compare particles arrival times under Gran Sasso,''
  arXiv:1208.0791 [hep-ex].
  N.~Y.~.Agafonova {\it et al.}  [LVD Collaboration],
  ``Measurement of the velocity of neutrinos from the CNGS beam with the Large Volume Detector,''
  Phys.\ Rev.\ Lett.\  {\bf 109} (2012) 070801
  [arXiv:1208.1392 [hep-ex]].
M.~Antonello, B.~Baibussinov, P.~Benetti, E.~Calligarich, N.~Canci, S.~Centro, A.~Cesana and K.~Cieslik {\it et al.},
  ``Precision measurement of the neutrino velocity with the ICARUS detector in the CNGS beam,''
  arXiv:1208.2629 [hep-ex].

\bibitem{k21987}
  K.~Hirata {\it et al.} [ KAMIOKANDE-II Collaboration ],
  ``Observation of a Neutrino Burst from the Supernova SN 1987a,''
  Phys.\ Rev.\ Lett.\  {\bf 58 } (1987)  1490-1493.

\bibitem{imb1987}
  R.~M.~Bionta, G.~Blewitt, C.~B.~Bratton, D.~Casper, A.~Ciocio, R.~Claus, B.~Cortez, M.~Crouch {\it et al.},
  ``Observation of a Neutrino Burst in Coincidence with Supernova SN 1987a in the Large Magellanic Cloud,''
  Phys.\ Rev.\ Lett.\  {\bf 58 } (1987)  1494.

\bibitem{strumia}
  G.~F.~Giudice, S.~Sibiryakov, A.~Strumia,
  ``Interpreting OPERA results on superluminal neutrino,''
  [arXiv:1109.5682 [hep-ph]].

\bibitem{gg}
  A.~G.~Cohen and S.~L.~Glashow,
  ``Pair Creation Constrains Superluminal Neutrino Propagation,''
  Phys.\ Rev.\ Lett.\  {\bf 107} (2011) 181803
  [arXiv:1109.6562 [hep-ph]].

\bibitem{xiaojun}
  X.~-J.~Bi, P.~-F.~Yin, Z.~-H.~Yu and Q.~Yuan,
  ``Constraints and tests of the OPERA superluminal neutrinos,''
  Phys.\ Rev.\ Lett.\  {\bf 107} (2011) 241802
  [arXiv:1109.6667 [hep-ph]].

\bibitem{sformati}
  G.~Amelino-Camelia, L.~Freidel, J.~Kowalski-Glikman and L.~Smolin,
  ``OPERA neutrinos and relativity,''
  arXiv:1110.0521 [hep-ph].   Y.~Ling,
  ``A note on superluminal neutrinos and deformed special relativity,''
  arXiv:1111.3716 [hep-ph].

\bibitem{alex}
  G.~Dvali, A.~Vikman,
  ``Price for Environmental Neutrino-Superluminality,''
   [arXiv:1109.5685 [hep-ph]].

\bibitem{fifth}
  L.~Iorio,
  ``Environmental fifth-force hypothesis for the OPERA superluminal neutrino phenomenology: constraints from orbital motions around the Earth,''
  [arXiv:1109.6249 [gr-qc]].

\bibitem{tedeschi}
 J.~Evslin,
  ``Challenges Confronting Superluminal Neutrino Models,''
Int J Mod Phys Conf Series: Proceedings of the Symposium on Cosmology and Particle Astrophysics (CosPA 2011) {\bf{10}} 159-168.
  arXiv:1111.0733 [hep-ph].
  A.~Hebecker and A.~Knochel,
  ``The Price of Neutrino Superluminality continues to rise,''
  arXiv:1111.6579 [hep-ph].

\bibitem{nuenergy}
P.~Gu, X.~Wang, X.~Zhang,
  ``Dark energy and neutrino mass limits from baryogenesis,''
  Phys.\ Rev.\  {\bf D68 } (2003)  087301.
  [hep-ph/0307148].

\bibitem{wiener}
  R.~Fardon, A.~E.~Nelson and N.~Weiner,
  ``Dark energy from mass varying neutrinos,''
  JCAP {\bf 0410} (2004) 005
  [astro-ph/0309800].

\bibitem{lsnd}
A.~Aguilar-Arevalo {\it et al.}  [LSND Collaboration],
  ``Evidence for neutrino oscillations from the observation of anti-neutrino(electron) appearance in a anti-neutrino(muon) beam,''
  Phys.\ Rev.\ D {\bf 64} (2001) 112007
  [hep-ex/0104049].

\bibitem{miniboone}
A.~A.~Aguilar-Arevalo {\it et al.}  [The MiniBooNE Collaboration],
  ``A Search for electron neutrino appearance at the $\Delta m^{2} \sim 1$eV$^{2}$ scale,''
  Phys.\ Rev.\ Lett.\  {\bf 98} (2007) 231801
  [arXiv:0704.1500 [hep-ex]].
A.~A.~Aguilar-Arevalo {\it et al.}  [MiniBooNE Collaboration],
  ``Unexplained Excess of Electron-Like Events From a 1-GeV Neutrino Beam,''
  Phys.\ Rev.\ Lett.\  {\bf 102} (2009) 101802
  [arXiv:0812.2243 [hep-ex]].
 A.~A.~Aguilar-Arevalo {\it et al.}  [The MiniBooNE Collaboration],
  ``Event Excess in the MiniBooNE Search for $\bar \nu_\mu \rightarrow \bar \nu_e$ Oscillations,''
  Phys.\ Rev.\ Lett.\  {\bf 105} (2010) 181801
  [arXiv:1007.1150 [hep-ex]].
 A.~A.~Aguilar-Arevalo {\it et al.}  [MiniBooNE Collaboration],
  ``A Combined $\nu_\mu \to \nu_e$ and $\bar\nu_\mu \to \bar\nu_e$ Oscillation Analysis of the MiniBooNE Excesses,''
  arXiv:1207.4809 [hep-ex].
 
\bibitem{karmen}
  B.~Armbruster {\it et al.}  [KARMEN Collaboration],
  ``Upper limits for neutrino oscillations muon-anti-neutrino ---> electron-anti-neutrino from muon decay at rest,''
  Phys.\ Rev.\ D {\bf 65} (2002) 112001
  [hep-ex/0203021].

\bibitem{icarus}
  M. Antonello, B.~Baibussinov, P.~Benetti, E.~Calligarich, N.~Canci, S.~Centro, A.~Cesana and K.~Cieslik {\it et al.},
  ``Experimental search for the LSND anomaly with the ICARUS LAr TPC detector in the CNGS beam,''
  arXiv:1209.0122 [hep-ex].

\bibitem{gallex}
P.~Anselmann {\it et al.}  [GALLEX. Collaboration],
  ``First results from the Cr-51 neutrino source experiment with the GALLEX detector,''
  Phys.\ Lett.\ B {\bf 342} (1995) 440.
W.~Hampel {\it et al.}  [GALLEX Collaboration],
  ``Final results of the Cr-51 neutrino source experiments in GALLEX,''
  Phys.\ Lett.\ B {\bf 420} (1998) 114.

\bibitem{sage}
D.~.N.~Abdurashitov, V.~N.~Gavrin, S.~V.~Girin, V.~V.~Gorbachev, T.~V.~Ibragimova, A.~V.~Kalikhov, N.~G.~Khairnasov and T.~V.~Knodel {\it et al.},
  ``The Russian-American gallium experiment (SAGE) Cr neutrino source measurement,''
  Phys.\ Rev.\ Lett.\  {\bf 77} (1996) 4708.
J.~N.~Abdurashitov {\it et al.}  [SAGE Collaboration],
  ``Measurement of the response of the Russian-American gallium experiment to neutrinos from a Cr-51 source,''
  Phys.\ Rev.\ C {\bf 59} (1999) 2246
  [hep-ph/9803418].
J.~N.~Abdurashitov, V.~N.~Gavrin, S.~V.~Girin, V.~V.~Gorbachev, P.~P.~Gurkina, T.~V.~Ibragimova, A.~V.~Kalikhov and N.~G.~Khairnasov {\it et al.},
  ``Measurement of the response of a Ga solar neutrino experiment to neutrinos from an Ar-37 source,''
  Phys.\ Rev.\ C {\bf 73} (2006) 045805
  [nucl-ex/0512041].
J.~N.~Abdurashitov {\it et al.}  [SAGE Collaboration],
  ``Measurement of the solar neutrino capture rate with gallium metal. III: Results for the 2002--2007 data-taking period,''
  Phys.\ Rev.\ C {\bf 80} (2009) 015807
  [arXiv:0901.2200 [nucl-ex]].

\bibitem{reattoreanom}
G.~Mention, M.~Fechner, T.~.Lasserre, T.~.A.~Mueller, D.~Lhuillier, M.~Cribier and A.~Letourneau,
  ``The Reactor Antineutrino Anomaly,''
  Phys.\ Rev.\ D {\bf 83} (2011) 073006
  [arXiv:1101.2755 [hep-ex]].

\bibitem{noiunokm}
 E.~Ciuffoli, J.~Evslin and H.~Li,
  ``The Reactor Anomaly after Daya Bay and RENO,''
  arXiv:1205.5499 [hep-ph].
 
\bibitem{smirnov}
P.~C.~de Holanda and A.~Y.~.Smirnov,
  ``Homestake result, sterile neutrinos and low-energy solar neutrino experiments,''
  Phys.\ Rev.\ D {\bf 69} (2004) 113002
  [hep-ph/0307266].
P.~C.~de Holanda and A.~Y.~.Smirnov,
  ``Solar neutrino spectrum, sterile neutrinos and additional radiation in the Universe,''
  Phys.\ Rev.\ D {\bf 83} (2011) 113011
  [arXiv:1012.5627 [hep-ph]].

\bibitem{spt}
  R.~Keisler, C.~L.~Reichardt, K.~A.~Aird, B.~A.~Benson, L.~E.~Bleem, J.~E.~Carlstrom, C.~L.~Chang and H.~M.~Cho {\it et al.},
  ``A Measurement of the Damping Tail of the Cosmic Microwave Background Power Spectrum with the South Pole Telescope,''
  Astrophys.\ J.\  {\bf 743} (2011) 28
  [arXiv:1105.3182 [astro-ph.CO]].

\bibitem{wmap7}
  E.~Komatsu {\it et al.}  [WMAP Collaboration],
  ``Seven-Year Wilkinson Microwave Anisotropy Probe (WMAP) Observations: Cosmological Interpretation,''
  Astrophys.\ J.\ Suppl.\  {\bf 192} (2011) 18
  [arXiv:1001.4538 [astro-ph.CO]].



%\bibitem{minos}
%  P.~Adamson {\it et al.} [ MINOS Collaboration ],
%  ``Measurement of neutrino velocity with the MINOS detectors and NuMI neutrino beam,''
%  Phys.\ Rev.\  {\bf D76 } (2007)  072005.
%  [arXiv:0706.0437 [hep-ex]].

\bibitem{icecube}
  R.~Abbasi {\it et al.} [ IceCube Collaboration ],
  ``Measurement of the atmospheric neutrino energy spectrum from 100 GeV to 400 TeV with IceCube,''
  Phys.\ Rev.\  {\bf D83 } (2011)  012001.
  [arXiv:1010.3980 [astro-ph.HE]].

\bibitem{nuenergyvec}
P.~-H.~Gu, X.~-J.~Bi, X.~-m.~Zhang,
  ``Dark energy and neutrino CPT violation,''
  Eur.\ Phys.\ J.\  {\bf C50 } (2007)  655-659.
  [hep-ph/0511027];
Shin'ichiro Ando, Marc Kamionkowski, Irina Mocioiu Neutrino
"Oscillations, Lorentz/CPT Violation, and Dark Energy" Phys.Rev.
{\bf D80}, (2009) 123522.

\bibitem{newerellis}
  J.~Alexandre, J.~Ellis, N.~E.~Mavromatos,
  ``On the Possibility of Superluminal Neutrino Propagation,''
  [arXiv:1109.6296 [hep-ph]].

\bibitem{grecia}
A.~Kehagias,
  ``Relativistic Superluminal Neutrinos,''
  [arXiv:1109.6312 [hep-ph]].

\bibitem{nano}
 T.~Li, D.~V.~Nanopoulos,
  ``Background Dependent Lorentz Violation from String Theory,''
  [arXiv:1110.0451 [hep-ph]].

\bibitem{Galileon}
  A.~Nicolis, R.~Rattazzi, E.~Trincherini,
  ``The Galileon as a local modification of gravity,''
  Phys.\ Rev.\  {\bf D79 } (2009)  064036.
  [arXiv:0811.2197 [hep-th]].

\bibitem{consistency}
  A.~Nicolis, R.~Rattazzi,
  ``Classical and quantum consistency of the DGP model,''
  JHEP {\bf 0406 } (2004)  059.
  [hep-th/0404159].


\bibitem{Vainshtein}
  A.~I.~Vainshtein,
  ``To the problem of nonvanishing gravitation mass,''
  Phys.\ Lett.\  {\bf B39 } (1972)  393-394.

\bibitem{taotao}
  J.~Evslin, T.~Qiu,
  ``Closed Timelike Curves in the Galileon Model,''
  [arXiv:1106.0570 [hep-th]].

\bibitem{kamland}
  S.~Abe {\it et al.} [ KamLAND Collaboration ],
  ``Precision Measurement of Neutrino Oscillation Parameters with KamLAND,''
  Phys.\ Rev.\ Lett.\  {\bf 100 } (2008)  221803.
  [arXiv:0801.4589 [hep-ex]].

\bibitem{msw}
 L.~Wolfenstein,
  ``Neutrino Oscillations in Matter,''
  Phys.\ Rev.\ D {\bf 17} (1978) 2369.
   S.~P.~Mikheev and A.~Y.~.Smirnov,
  ``Resonance Amplification of Oscillations in Matter and Spectroscopy of Solar Neutrinos,''
  Sov.\ J.\ Nucl.\ Phys.\  {\bf 42} (1985) 913
   [Yad.\ Fiz.\  {\bf 42} (1985) 1441].

\bibitem{strumiarev}
  A.~Strumia and F.~Vissani,
  ``Neutrino masses and mixings and...,''
  hep-ph/0606054.

\bibitem{stars}
K.~Nomoto,
 ``Evolution of 8 to 10 solar mass stars toward electron capture supernovae II,"
Astro.\ Jour.\ {\bf 322} (1987) 206.

\bibitem{LEP}
  B.~Altschul,
  ``Bounding Isotropic Lorentz Violation Using Synchrotron Losses at LEP,''
  Phys.\ Rev.\  {\bf D80 } (2009)  091901.
  [arXiv:0905.4346 [hep-ph]].

\end{thebibliography}
\end{document}